\documentclass[twocolumn]{aastex63}
\usepackage{lineno}

\shorttitle{An Improved GPU-Based Ray-Shooting Code For Gravitational Microlensing}
\shortauthors{Wenwen Zheng et al.}

\graphicspath{{./}{figures/}}

\usepackage{amsmath}	
\usepackage{amssymb}

\begin{document}
\begin{sloppypar}

\title{An Improved GPU-Based Ray-Shooting Code For Gravitational Microlensing}

\email{Contact e-mail: guoliang@pmo.ac.cn (G. Li);}
\author[0000-0002-0478-3431]{Wenwen Zheng}
\affil{Purple Mountain Observatory, Chinese Academy of Sciences, Nanjing, Jiangsu, 210023, China}
\affil{School of Astronomy and Space Science, University of Science and Technology of China, Hefei, Anhui, 230026, China}

\author{Xuechun Chen}
\affil{Purple Mountain Observatory, Chinese Academy of Sciences, Nanjing, Jiangsu, 210023, China}
\affil{School of Astronomy and Space Science, University of Science and Technology of China, Hefei, Anhui, 230026, China}
\author{Guoliang Li}
\affil{Purple Mountain Observatory, Chinese Academy of Sciences, Nanjing, Jiangsu, 210023, China}
\author{Hou-zun Chen}
\affil{Purple Mountain Observatory, Chinese Academy of Sciences, Nanjing, Jiangsu, 210023, China}
\affil{School of Astronomy and Space Science, University of Science and Technology of China, Hefei, Anhui, 230026, China}



\begin{abstract}

We present an improved inverse ray-shooting code based on GPUs for generating microlensing magnification maps. 
In addition to introducing GPUs for acceleration, we put the efforts in two aspects: (i) A standard circular lens plane is replaced by a rectangular one to reduce the number of unnecessary lenses as a result of an extremely prolate rectangular image plane. (ii) Interpolation method is applied in our implementation which has achieved an significant acceleration when dealing with large number of lenses and light rays required by high resolution maps.
With these applications, we have greatly reduced the running time while maintaining high accuracy: the speed has been increased by about 100 times compared with ordinary GPU based IRS code and GPU-D code when handling large number of lenses. If encountered the high resolution situation up to $10000^2$ pixels, resulting in almost $10^{11}$ light rays, the running time can also be reduced by two orders of magnitude.

\end{abstract}

\keywords{Gravitational lensing -- Microlensing -- Methods:Numerical}

\section{Introduction} \label{sec:intro}

Gravitational lensing studies on the source light deflected by the foreground lens has become a booming field since the first discovery of multi-imaged quasar \cite{1979a279..381WtNur.}.
In the meanwhile, \cite{1979Natur.282..561C} realized that the compact mass distributions within lens galaxies may further impact the light and split the image into sub-images, whose angular separation is too small to resolve, naming cosmological microlensing effect.
A decade later, an unambiguous evidence for a brightness change of one image of quasar system Q2237+0305 was presented by \cite{1989AJ.....98.1989I}, representing the first detected microlensing event.
Since then, microlensing has been developed as a powerful tool in several aspects, such as probing inner structures of quasars, e.g. accretion disk \cite{2005ApJ...628..594M}, \cite{2010ApJ...712.1129M}, \cite{2016AN....337..356C} and quasar broad line region \cite{2012A&A...544A..62S}, \cite{2013ApJ...764..160G}, \cite{2018ApJ...859...50F}, constraining the compact matter distribution of lens galaxies \cite{2000MNRAS.315...51W}, \cite{2001MNRAS.320...21W}, \cite{2007MNRAS.376..263C}, \cite{2009ApJ...706.1451M}, \cite{2017ApJ...836L..18M}, \cite{2019ApJ...885...75J}.

The essential part of microlensing study is generating magnification maps of the source plane. The standard method for producing microlensing magnification patterns is the inverse ray-shooting (IRS) which is first described by \cite{1986A&A...166...36K}, \cite{1986MPARp.234.....S}. One shoots large number of light rays back from observer (image plane) to the source plane, then calculates the deflection angle caused by hundreds to millions of micro-lenses.
However, the brute-force IRS contains a great number of operations, which is proportional to the number of light rays ($N_{rays}$) and microlenses ($N_*$), resulting in catastrophic time consuming when millions of lenses are required in parameter space while obtaining a magnification map with satisfied resolution and accuracy.

Great efforts have been made to deal with this problem towards two directions: (i) To parallel the computing. (ii) To reduce the number of lenses ($N_*$) or light rays ($N_{rays}$) involved in the calculations.
For the first direction, different parallel strategies based on both CPUs \cite{2017A&C....19...60C} and GPUs have been applied on IRS method, among which GPUs have shown great computational potential \cite{2010NewA...15...16T}.
For the second direction, \cite{1999JCoAM.109..353W} introduced hierarchical tree method to reorganize the lenses by their distance to individual light and group distant lenses into larger cells, allowing $N_*$ directly involved in calculations to be reduced. While the IPM method described in \cite{2006ApJ...653..942M}, \cite{2011ApJ...741...42M} suggests mapping regular polygons instead of individual light, so that the magnification is related to the ratio of the overlapped area of mapped polygon and source pixel to the original area. With this approach, the required number of $N_{rays}$ is greatly reduced to maintain the same accuracy.

Improved algorithms are still coming up, \cite{2010NewA...15..181G} implemented a CPU-based tree code on supercomputer.
\cite{Teralens} presented a GPU parallel tree-code . \cite{2021arXiv210508690W} introduced another GPU-based code which calculates the Taylor coefficients of the four corners of Sparse squares and shoots more rays within the squares to use those coefficients for acceleration, all lenses are used in the meantime.
\cite{2021arXiv210707222S} brought up a new non-parallel method which combine the IPM algorithm with a Possion solver for a deflection potential to reduce the $N_*$ and $N_{rays}$ simultaneously.

In this work, we present a GPU-based inverse ray-shooting code implemented with a rectangular lens plane and combined with the interpolation algorithm to improve the efficiency. We name it GPU-PMO (GPU-Parallel Microlensing Optimizer) hereafter.
This paper is organized as follow: section~\ref{sec:Methods} outlines the main procedures of our work, especially focusing on the modified rectangular lens plane, the setting of the ``protective border'' and the interpolation method.
Section~\ref{sec:Results} presents the run-time and accuracy comparison between our GPU-PMO code, the ordinary GPU-based IRS code and the widely used GPU-D code.
In section~\ref{sec:Summary}, we provide a short summary of this work.

\section{Methods} \label{sec:Methods}
\subsection{Lens Equation} \label{subsec:Lens Equation}
The core idea of IRS method is to select a small rectangular ``window'' on the image plane (usually within a larger circular lens plane), and trace back a large number of light rays from the image window to the source plane based on lens equation:
\begin{equation}
\scriptsize
\pmb\beta = \left( 
\begin{array}{cc}
1 - (\kappa-\kappa_*)-\gamma & 0 \\
0 & 1 - (\kappa-\kappa_*)+\gamma 
\end{array}
\right) {\pmb\theta}
- \sum_{i=1}^{N_*} m_i \frac{({\pmb \theta} 
- {\pmb \theta}_i ) }
{\vert{\pmb \theta} - {\pmb \theta}_i\vert^2},
\label{eqn:lensequation}
\end{equation}
where $\pmb\beta$ refers to the light positions in the source plane, $\pmb\theta$ and $\pmb\theta_i$ are respectively the light positions and lens positions on the image/lens plane. $N_*$ stands for the number of microlenses while $m_i$ denotes their mass. In addition, $\kappa$ represents the dimensionless surface mass density of the mass sheet of the lens plane and $\gamma$ represents the external shear, both can be obtained from the strong lens model. 

Since the lens plane can be regard as a mixture of continuously distributed matter (generally considered to be the dark matter) and compact objects such as stars and black holes, $\kappa$ can be divided into two parts: smooth part $\kappa_s$ and compact  part $\kappa_*$, i.e.,
\begin{equation}
\kappa=\kappa_s+\kappa_*.
\end{equation}

The lens Eq.~(\ref{eqn:lensequation}) is adopted with angular coordinates and scaled by Einstein angle $\theta_E$ of a point mass:
\begin{equation}
    \theta_E=\sqrt{{\frac{4G\left\langle M\right\rangle}{c^2}}{\frac{D_{ls}}{D_lD_s}}},
	\label{eq:thetaE}
\end{equation}
where $\left\langle M\right\rangle$ stands for the average mass of the micro-lenses, $D_{ls}$, $D_l$ and $D_s$ refers to the angular diameter distances from the lens to the source, from the lens to the observer, and from the source to the observer, respectively.

If Eq.~(\ref{eqn:lensequation}) is rewritten in the form of:
\begin{equation}
\pmb \beta=\pmb \theta-\pmb \alpha\left(\pmb \theta\right),
\end{equation}
we obtain the equation of deflection angle:
\begin{equation}
{\pmb \alpha\left(\pmb \theta\right)} = \left( 
\begin{array}{cc}
\kappa_s+\gamma & 0 \\
0 & \kappa_s-\gamma 
\end{array}
\right) {\pmb \theta}
+\sum_{i=1}^{N_*} m_i \frac{\left({\pmb \theta} - {\pmb \theta}_i \right)}
{\left|\pmb \theta - \pmb \theta_i\right|^2},
\label{eqn:alpha}
\end{equation}
which clearly reveals the essence of microlensing: the light rays will be affected by both the smooth matter and compact objects (micro-lenses) in the path, resulting in the deflection angles.
The first term of Eq.~(\ref{eqn:alpha}) represents the contribution from smooth matter and external shear, the second term refers to the cumulative contribution to the deflection angle from each micro-lense.
The main procedure of IRS method is to solve the deflection angle of each light ray, which is the most time-consuming part. The bright side is, the calculation of each light ray is relatively independent, which is perfectly suitable for GPU parallel processing.

\subsection{Modified rectangular lens plane}
\begin{figure}
	\includegraphics[width=\columnwidth]{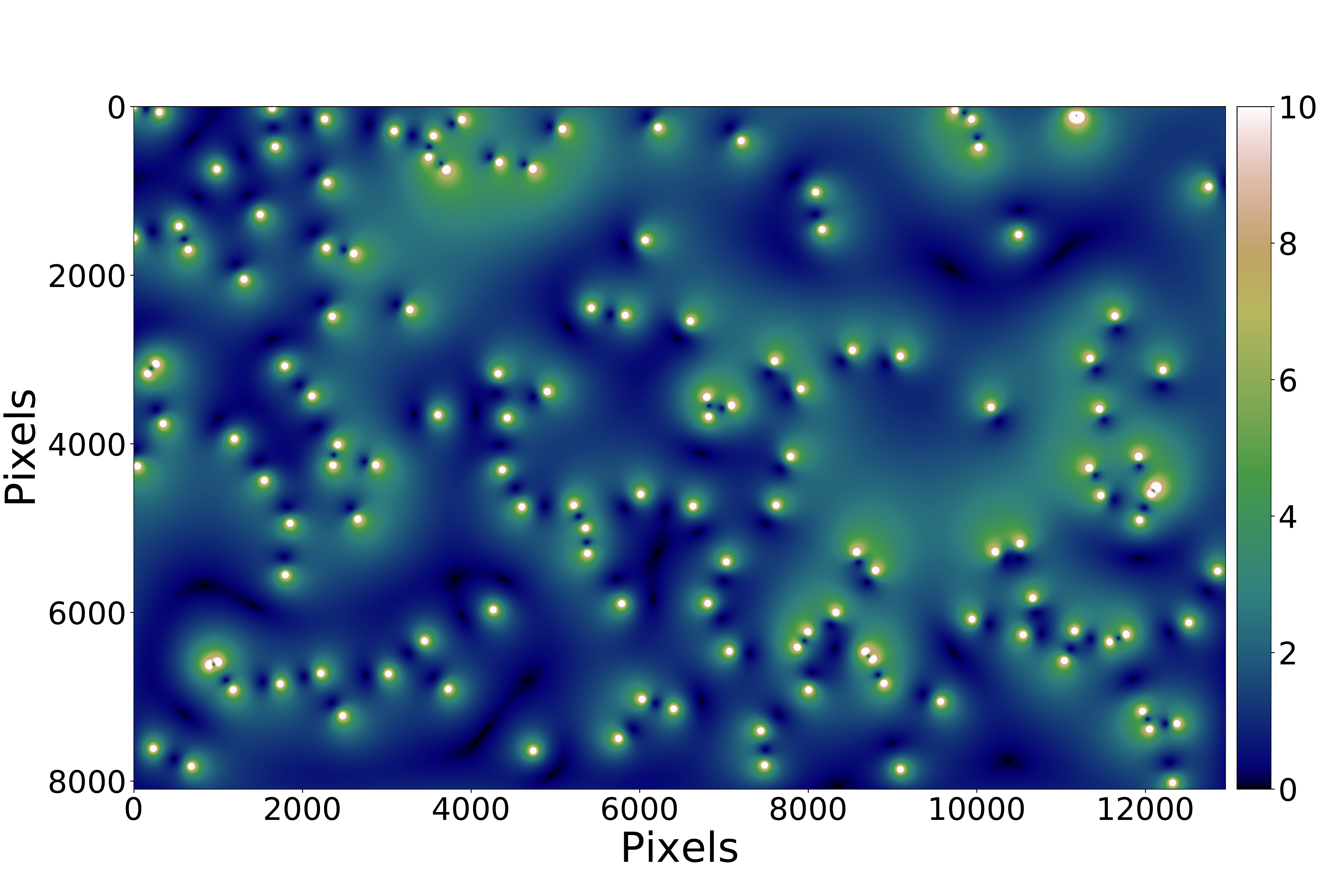}
	
    \caption{Deflection angle perturbation in lens plane. $\kappa=\kappa_*=0.6$, $\gamma=0.1$, $N_*=238$, with equal mass of 1$M_\odot$.
            }
    \label{fig:Alpha}
    
\end{figure}
In our implementation, we choose a rectangular plane acts both as an image 
plane and a lens plane, as shown in Fig.~\ref{fig:Threelevels}, rather than a rectangular image plane embedded in a larger circular lens plane. 
Changing a circular lens plane into a rectangular one is not just changing the shape, but also the lens equation.

Notably, neglecting the shape of the lens plane, the general form of deflection angle can be written as:
\begin{equation}
{\pmb \alpha(\pmb \theta)} =  \left( 
\begin{array}{cc}
\kappa+\gamma & 0 \\
0 & \kappa-\gamma 
\end{array}
\right) {\pmb \theta}
+\sum_{i=1}^{N_*} m_i \frac{({\pmb \theta} - {\pmb \theta}_i ) }
{\vert{\pmb \theta} - {\pmb \theta}_i\vert^2}
+\pmb \alpha_{-\kappa_*},
\label{eqn:alpha1}
\end{equation}
the first term refers to the influence of local convergence ($\kappa$, from both the compact and smooth mass) and shear ($\gamma$) caused by strong lensing .
The second term indicates the contribution from the individual micro-lenses, whose  mean surface mass density is $\kappa_*$.
One may find that the mass sheet with density of $\kappa_*$ is redundant, thus we add the third term $\pmb\alpha_{-\kappa_*}$ which represents a negative mass sheet that contains only smooth  matter with the same $\kappa_*$ to cancel out the ``extra mass'' from the second term.
The sum of the second and the third term can be regard as the perturbation caused by compact objects.

The specific form of the third term, which also consists of a ``convergence part'' and a ``shear part'', is associated with the shape of the lens plane.
\begin{figure*}
	\includegraphics[width=\linewidth]{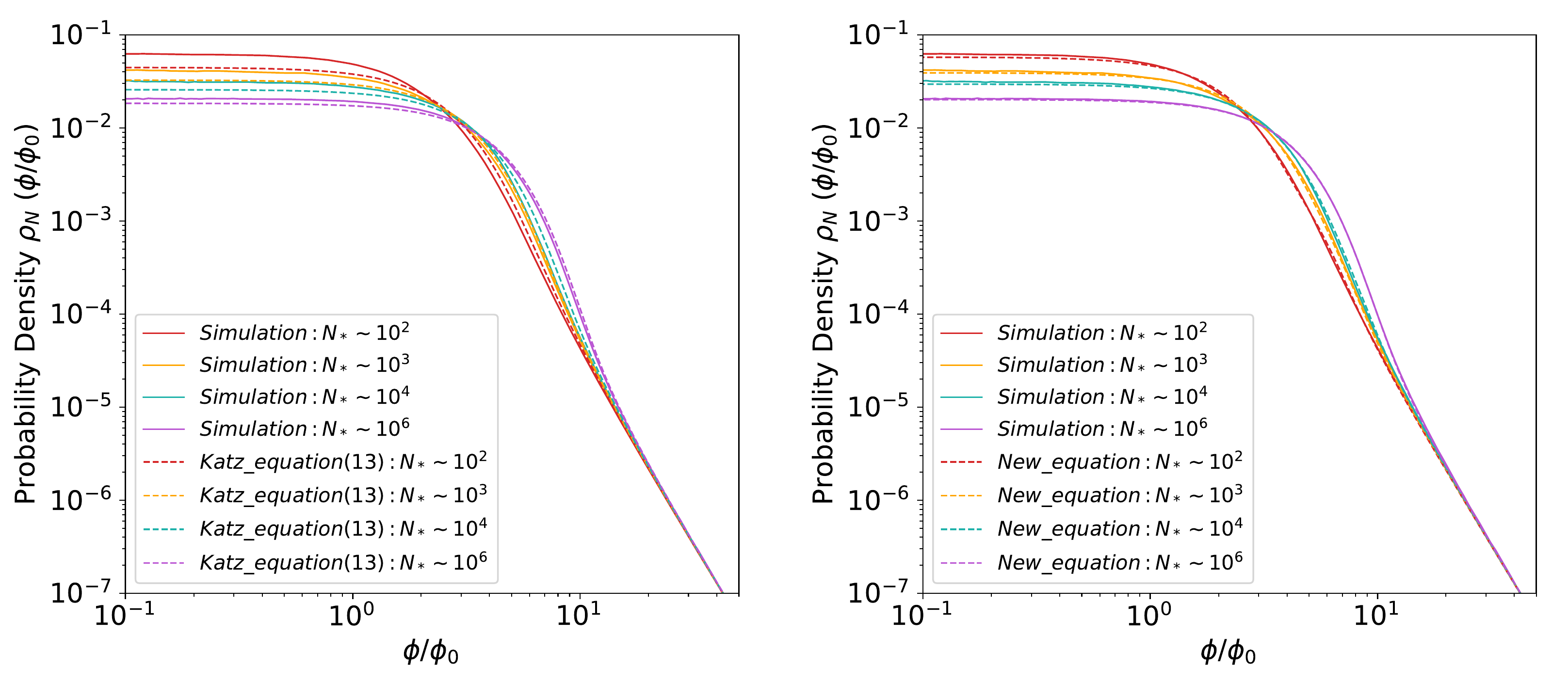}
    \caption{The probability density distribution of deflection angle perturbation with different number of lenses, solid lines represent the deflection angle distribution of our numerical simulation, dashed lines represent the analytical results of equation (13) from the \cite{1986ApJ...306....2K} (left) and analytical results from our modified equation (right),respectively. different colors stands for deflection angle distributions with lenses of $10^2$, $10^3$, $10^4$, $10^6$. 
    }
    \label{fig:Proscale}
\end{figure*}
For a circular lens plane (mass sheet), the deflection angle is given by $\pmb\alpha$ $\propto$ M/$\pmb\theta$ = -$\kappa_*$$\pi$$\pmb\theta$, which is a combination of two vector components: $\alpha_1$ $\propto$ -$\kappa_*$$\pi$$\theta_1$, $\alpha_2$ $\propto$ -$\kappa_*$$\pi$$\theta_2$, here M and $\pmb\theta$ denote the mass inside an angular radius of $\pmb\theta$. Accordingly, the ``shear'' part would be:
\begin{equation}
    \gamma_1=\frac{1}{2}(\frac{\partial \alpha_1}{\partial \theta_1}-\frac{\partial \alpha_2}{\partial \theta_2})
    =0,
	\label{eq:shear1}
\end{equation}
\begin{equation}
    \gamma_2=\frac{\partial \alpha_2}{\partial \theta_1}-\frac{\partial \alpha_1}{\partial \theta_2}
    =0,
	\label{eq:shear2}
\end{equation}
therefore, for a circular lens plane, $\pmb\alpha_{-\kappa_*}$ contains only the part of  ``convergence'':
\begin{equation}
{\pmb \alpha_{-\kappa_*}} =  \left( 
\begin{array}{cc}
-\kappa_* & 0 \\
0 & -\kappa_* 
\end{array}
\right) {\pmb \theta}.
\label{eqn:alphapart}
\end{equation}
\\
Then we deduce Eq.~(\ref{eqn:alpha}) by combining Eq.~(\ref{eqn:alphapart}) and Eq.~(\ref{eqn:alpha1}).
It is worth noting that Eq.~(\ref{eqn:lensequation}) and Eq.~(\ref{eqn:alpha}) only default to the circular lens plane which is commonly used, for lens planes with other shape such as the rectangular one, the ``shear'' part of $\pmb\alpha_{-\kappa_*}$ is no longer zero.

In this work, we obtained the modified $\pmb\alpha_{-\kappa_*}$ suit for a rectangular lens plane, please see Appendix for more details.
An example of the deflection angle perturbation for $\kappa_*\ne 0$ is shown in Fig.~\ref{fig:Alpha}, which is the sum of the second (compact mass) and the third term (negative mass sheet  $\pmb\alpha_{-\kappa_*}$) of Eq.~(\ref{eqn:alpha1}).

The following question to address might be: why bother using a rectangular lens plane rather than a circular one? As we know that Eq.~(\ref{eqn:lensequation}) requires a circular lens plane which is the circumscribed circle of the image rectangle with a side length ratio of $\mid1-\kappa-\gamma\mid$ / $\mid1-\kappa+\gamma\mid$.
The area of lens circle and image rectangle will not make a significant difference when $\mid\gamma\mid$ $\ll$ $\mid1-\kappa\mid$.
However, the area of the circumscribed circle will be dramatically larger than its inscribed rectangle when the image plane is extremely long and narrow, and the circumscribed circle will be filled with lots of ``wasted lenses'', result in excess calculations.
The modified lens plane shares the same rectangle with the image plane, solves the problem well, and it leads an easy way to set the ``protective border'' upon the source plane and corresponding image plane as shown below.

\subsection{The `` Protective Border ''}

Another problem comes afterwards: light rays from the image plane will not completely fall on the target source plane due to the compact objects in lens galaxies, which leads to the loss of magnification accuracy \cite{1986ApJ...306....2K},\cite{1999JCoAM.109..353W}.
Our approach is to add an ``protective border'' upon the target source plane as mentioned by \cite{1986ApJ...306....2K}, resulting in the enlarged corresponding image/lens plane to ensure that 98\% of the light rays fall on the target source plane to maintain accuracy. 
We follow \cite{1986ApJ...306....2K} to determine the size of the ``protected border'': we calculate the probability density distribution of deflection angle perturbation (as shown in Fig.~\ref{fig:Alpha}), but with varying number of lenses, presented in Fig.~\ref{fig:Proscale}.
Deflection angle $\alpha$ is scaled by $\phi_0$, here $\phi_0=\sqrt{\kappa_*}$, different colors denote different number of lenses from $10^2$ to $10^7$. Here we only show $10^2$, $10^3$, $10^4$ and $10^6$ in Fig.~\ref{fig:Proscale} for clarity. Solid lines represent the probability density calculated by our realization, dashed lines in the left panel represent the analytical results in Fig.1 of  \cite{1986ApJ...306....2K}, based on their Eq.(13) :
\begin{equation}\small
 \rho_N(\phi) = \frac{1}{2\pi\phi_0^2} \int_{0}^{\infty} x\mathrm{d}x\mathrm{J}_0\left ( \frac{x\phi}{\phi_0} \right ) \exp\left [ - \frac{x^2}{2}\ln_{}{\left (\frac{3.05N^{1/2}}{x}\right )}\right ].
 \label{eqn:Katz}
\end{equation}
Considering the term of  $\pmb \alpha_{-\kappa_*}$ of Eq.~(\ref{eqn:alpha1}) we find an improved Fourier transform of $\rho_N$:
\begin{equation}
\tilde\rho_N\simeq \exp\left [ -\frac{Nk^2}{2}\ln_{}{\frac{1.454}{k} }\right ],
\end{equation}
the form is consist with Katz Eq.(9) but with a changed numerator from 3.05 to 1.454. Thus we re-plot Eq.~(\ref{eqn:Katz}) with this change in the right panel of Fig.~\ref{fig:Proscale}.
Please refer to our follow-up work for more details.

As shown in Fig.~\ref{fig:Proscale}, our modified equation fits the numerical results better compared with the Katz Eq.(13). 
The broadening of the probability density distribution increases with the number of lenses, even for $10^7$ lenses. However, the distribution curves decay rapidly when deflection angles are larger than $4\phi_0$, the density is below $10^{-4}$ at $10\phi_0$.
As a result, we add a ``protective border'' with the length of $10\phi_0$ ($20\phi_0$ for side length in each direction) upon the target source plane, then the size of the corresponding image plane will be calculated based on this ``enlarged'' source plane. This process can ensure that more than 98$\%$ of the total required light rays fall into the target source plane and achieve a guaranteed accuracy.

\subsection{Interpolation algorithm}
As mentioned above, most calculating time is cost on computing deflection angles for individual light. Using GPU to parallel the computing of light rays just opens the outer loop, however, for each light ray, the inner loop that adds up the contribution of each lens to the deflection angle would significantly slow down the speed, especially when the number of lenses is getting larger \cite{2010NewA...15..726B}.

In this work, we quote the interpolation algorithm which is a tree-like method \cite{1999JCoAM.109..353W} to improve the computational efficiency. 
It is mainly about treating lenses differently by their distance to the individual light rays. 
We divide the lenses into distant ones and nearby ones, since the farther a lens is from the light ray, the less influence it brings. Thus for ``distant lenses'', we use interpolation to approximate their contribution to deflection angle, instead of simply summing up the lenses.

\begin{figure}
	\includegraphics[width=\columnwidth]{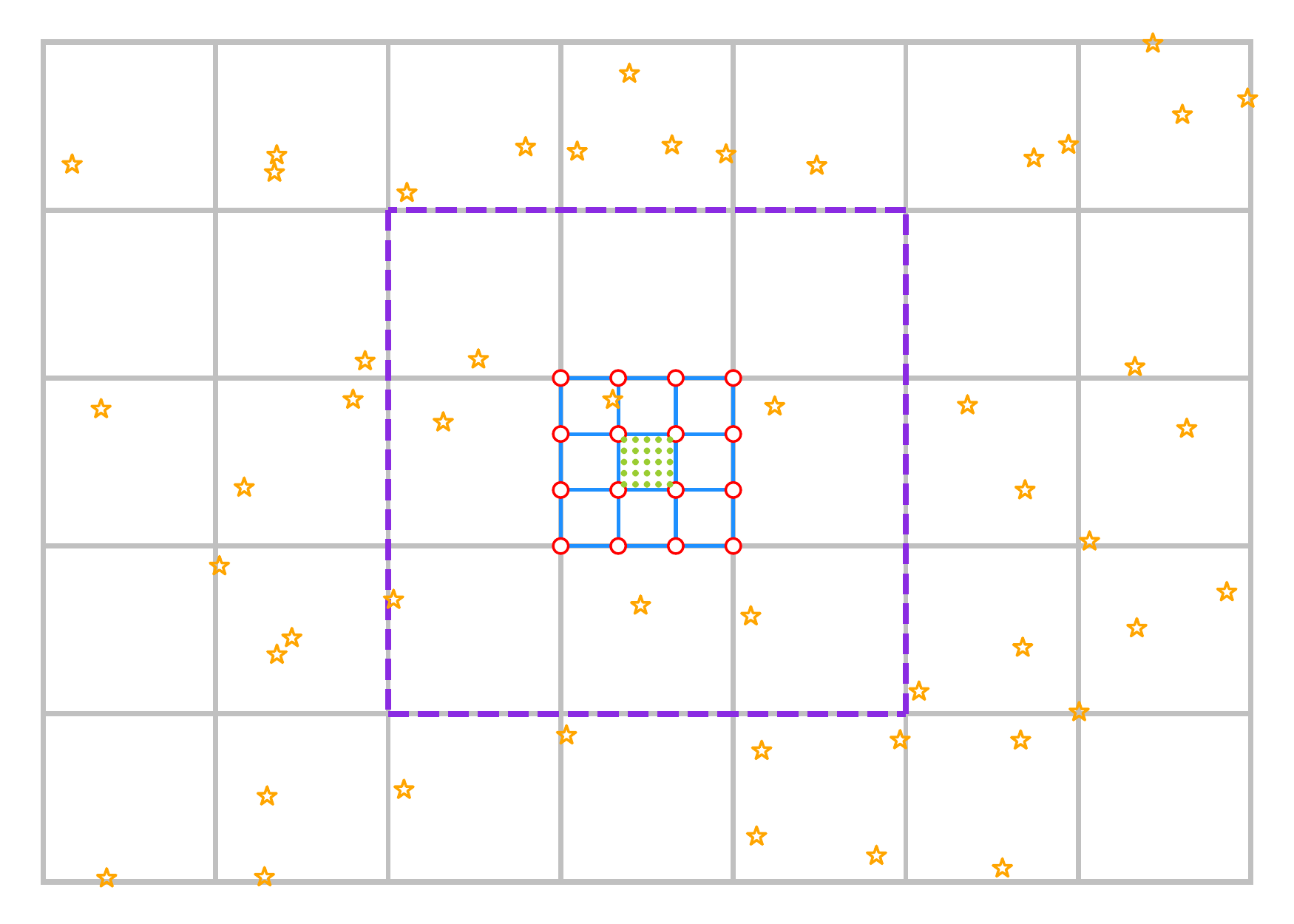}
    \caption{This figure shows a rectangular image/lens plane, which is organized in grey level-1 grid, yellow stars stand for randomly generated lenses. Green dots in the centered level-1 grid are a beam of sampled light rays. Blue grids represent the level-2 grids, red circles are level-2 grid points. 
    Purple dashed line is set to divide the ``distant lenses'' and ``nearby lenses'' for target light rays.
    }
    \label{fig:Threelevels}
\end{figure}

We achieve this by setting up three-level of grids in the image/lens plane, as shown in Fig.~\ref{fig:Threelevels}.
The grey rectangle denotes the image/lens plane, which is organized in grey level-1 grid to pack up the lenses (yellow stars). The level-1 grid in the center is chosen to be an example containing a beam of sampled light rays (green points, also called the level-3 grid points, in practice, the image plane will be filled with these light rays).
This central level-1 grid is further divided into blue level-2 grids (level-2 grid points are marked in red), and eight level-1 grids around it are marked out by purple dotted lines. 
For target light rays, the lenses inside the purple dashed lines will be treated as the ``nearby lenses'' and will be accumulated precisely. By contrast, the outside ones will be regard as the ``distant lenses'' and will be handled with the following interpolation procedures. The deflection angle will be the sum of the contributions from these two parts. 

For ``distant lenses'', their contribution will be accumulated upon the red level-2 grid points (for further optimization, the contribution from negative mass sheet will also be calculated in this step).
Then eight Taylor coefficients up to fourth order of level-2 grid points can be calculated directly, similar to the Appendix A.2 from \cite{1990PhDT.......180W}. Thus the deflection angle of target light rays contributed from ``distant lenses'' and the negative mass sheet can be easily obtained from the local Taylor expansion during the calculation of level-3 grids.
\\
\\
\begin{figure}
	\centering\includegraphics[width=7cm]{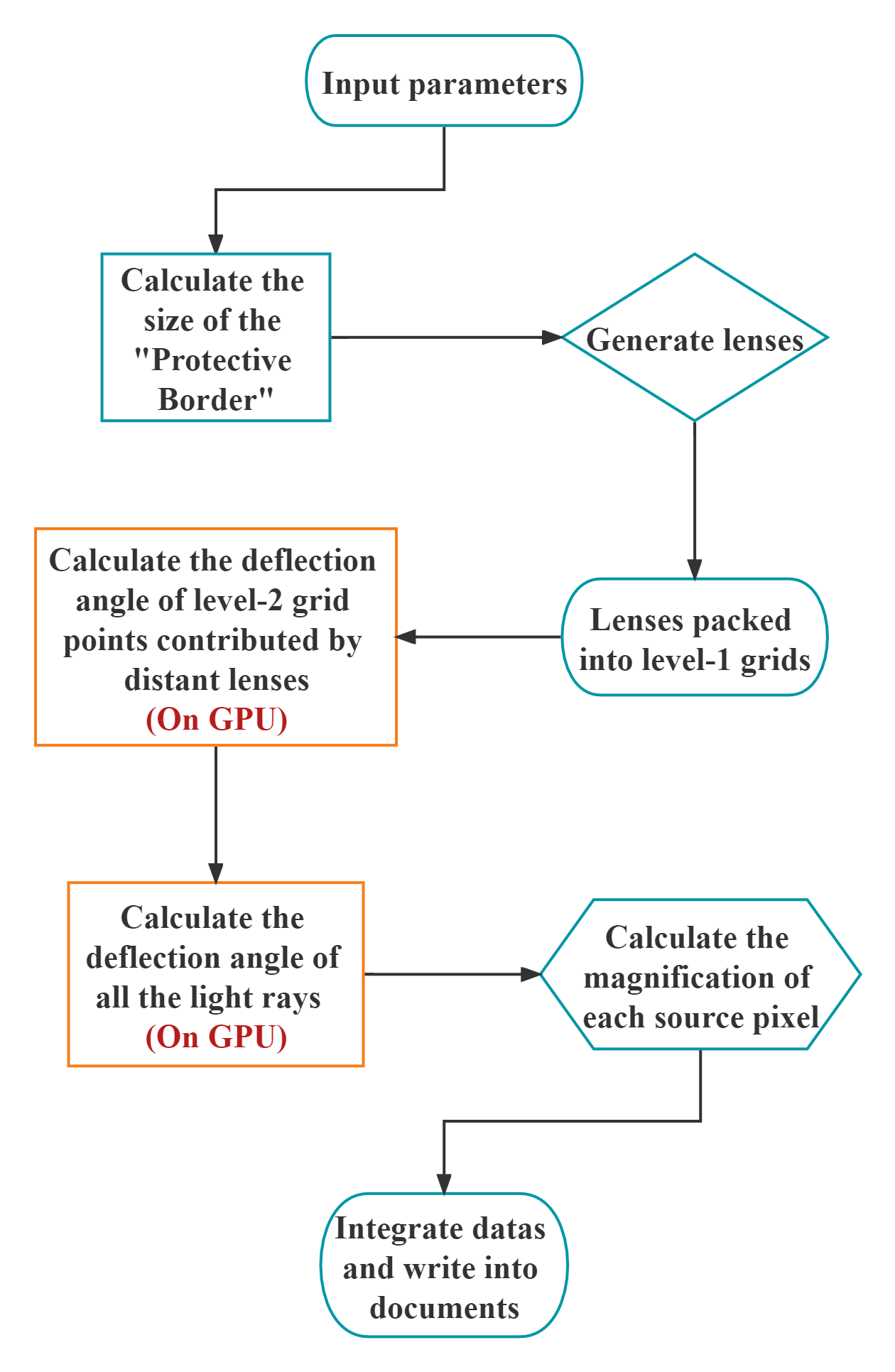}
    \caption{Flow chart of GPU-PMO code.
    }
    \label{fig:Flow chart}
\end{figure}
Fig.~\ref{fig:Flow chart} shows the flow chart of GPU-PMO code, i.e.
\begin{itemize}
\setlength{\itemindent}{0.0em}
\setlength{\leftmargin}{0.5em}
\setlength{\itemsep}{1.0ex} 
\setlength{\labelsep}{0.5em} 
\setlength{\listparindent}{0.0em} 
    \item \pmb{Input parameters} \\
          $\kappa$, $\gamma$, $f_*$: obtained from strong lens model, $f_* = \kappa_*/\kappa$;\\
          S\_scale (side length of source plane, scaled with $\theta_E$), $N_{pix}$( number of source pixels ), $N_{av}$ (average light rays per pixel): related to specific scientific requirements;\\
          $M_{min}$, $M_{max}$, $\alpha$: parameters of microlens mass when following certain mass distribution.
    \item \pmb{Calculate the size of the ``Protective Border''}\\
          P\_scale = 10$\sqrt{\kappa_*}$,\\
          then the side length of image/lens plane will be:\\
          I\_xscale = (S\_scale+2P\_scale) / $\vert{1-\kappa-\gamma}\vert$,\\
          I\_yscale = (S\_scale+2P\_scale) / $\vert{1-\kappa+\gamma}\vert$.
    \item \pmb{Generate lenses}\\
          Generate random positions and mass for lenses following certain mass distributions.
    \item \pmb{Lenses packed into level-1 grids}\\ 
          The resolution of level-1 grid:\\
          $\Delta x_1$= min\{$L_0$, min\{I\_xscale, I\_yscale\}/10\},\\
          where $L_0$=$\sqrt{(\mathrm{I\_xscale}\cdot \mathrm{I\_yscale})/N_*}$  represents the mean side length of the area occupied by each lens. We choose the smaller one from $L_0$ and 1/10 of the short side length.
    \item \pmb{Calculate the deflection angle of level2 grid points}\\ 
          The contribution of distant lenses and negative mass sheet ($\pmb\alpha_{-\kappa_*}$, contains complex calculations) are calculated in this step.\\
          The resolution of level-2 grid:\\
          $\Delta x_2= \Delta x_1$/20,\\
          apparently, the interpolation accuracy will be reduced if level-2 grid is too sparse, on the other hand, it's not worthy at the expense of speed to deal with the dense grid, after a series of tests we choose $\Delta x_1$/20 as a trade off.\\
          This step is implemented on GPU.
    \item \pmb{Calculate the deflection angle of all the light}\\
          \pmb{rays (level-3 grids)}\\
          The deflection angle consists of three parts, as described in Eq.~(\ref{eqn:alpha1}), which will be summed up in this step.
          The second term (i.e. the contribution from the individual micro-lenses) is the sum of the contribution from ``nearby lenses'' obtained from direct accumulation, and the “distant lenses” obtained from the local Taylor expansion of level-2 grid points.
          The contribution from the third term (i.e. the negative mass sheet $\pmb\alpha_{-\kappa_*}$) is calculated together with the ``distant lenses''.
          \\
          The resolution of level-3 grid, which is also the resolution of image plane:\\
          $\Delta x_3$ = $\sqrt{(\mathrm{I\_xscale} \cdot \mathrm{I\_yscale})/N_{rays}}$,\\
          where \, $N_{rays}$ = $N_{av}$ $\cdot$ $N_{pix}$, refers to the total number of shooting light rays.\\
          This step is implemented on GPU.
    \item \pmb{Calculate the magnification of each pixel}\\
             $\mathrm{Mag}_{ij} = N_{ij} \cdot  \frac{S_I}{S_S}$,\\
          where $N_{ij}$ represents the number of light rays fell into each source pixel,  $S_I$ and $S_S$ represents the pixel area of image plane and the source plane, respectively.
    \item \pmb{Integrate datas and write into the documents}
\end{itemize}



\begin{figure}
	\includegraphics[width=\columnwidth]{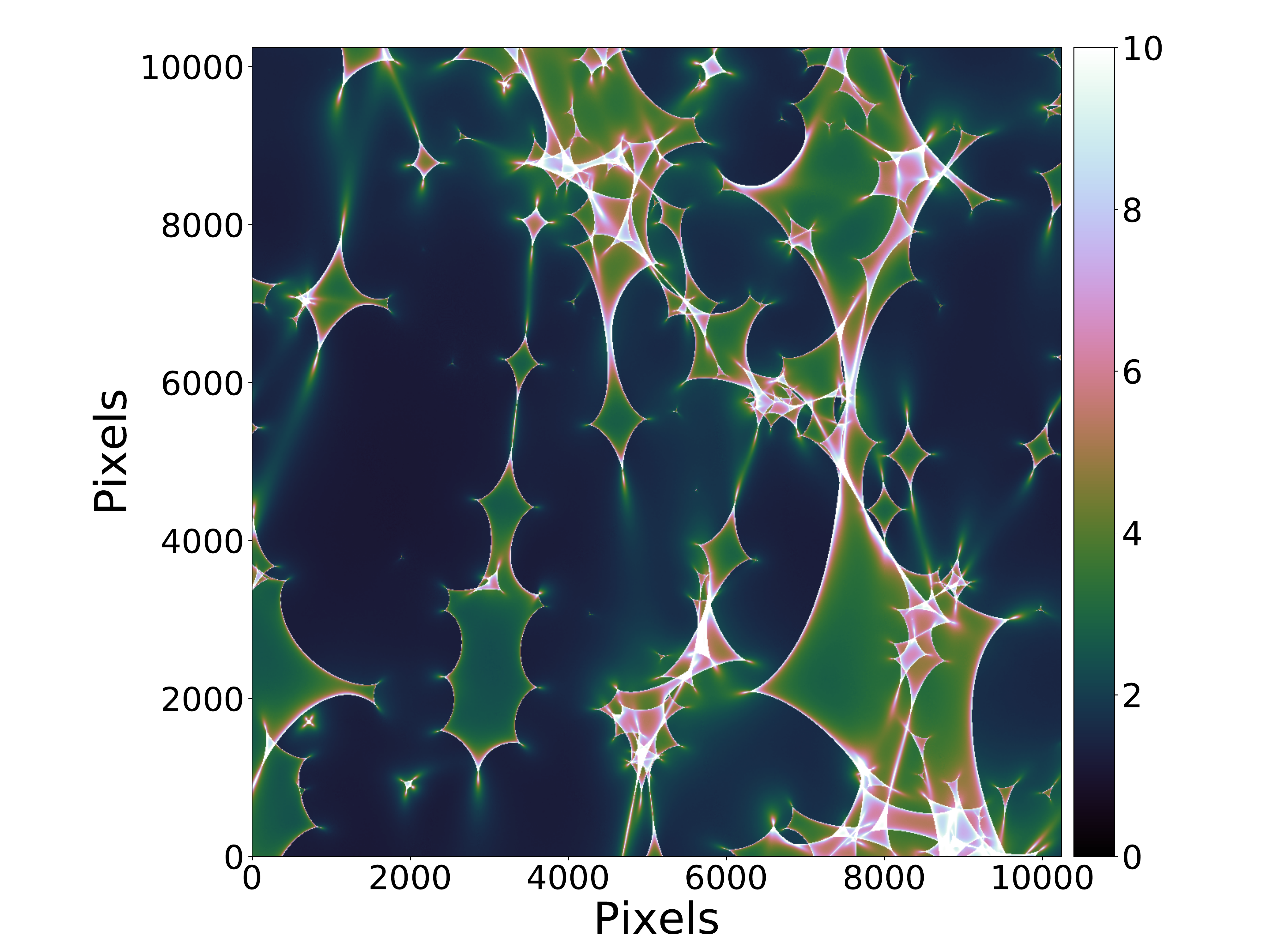}
    \caption{A sample of magnification map with a side-length of 20$\theta_E$ and a resolution of $10240^2$ pixels. Basic parameters: $\kappa$=0.3, $\gamma$=0.3, $f_*$=1.0, Nav=1000, a Salpeter mass distribution with $\left \langle M \right \rangle =0.3M_{\odot}$, corresponding to $N_*$= 238.
    }
    \label{fig:Performance}
\end{figure}

\section{Results}\label{sec:Results}

\;\;\;A sample of magnification map which is generated with our GPU-PMO code executed by an Nvidia Tesla V100S GPU is shown in Fig.~\ref{fig:Performance}.
To evaluate the performance of GPU-PMO code, we run a series of timing and accuracy tests and compare it with the mature GPU-D code.

\subsection{Speed}
\begin{figure}
	\includegraphics[width=\columnwidth]{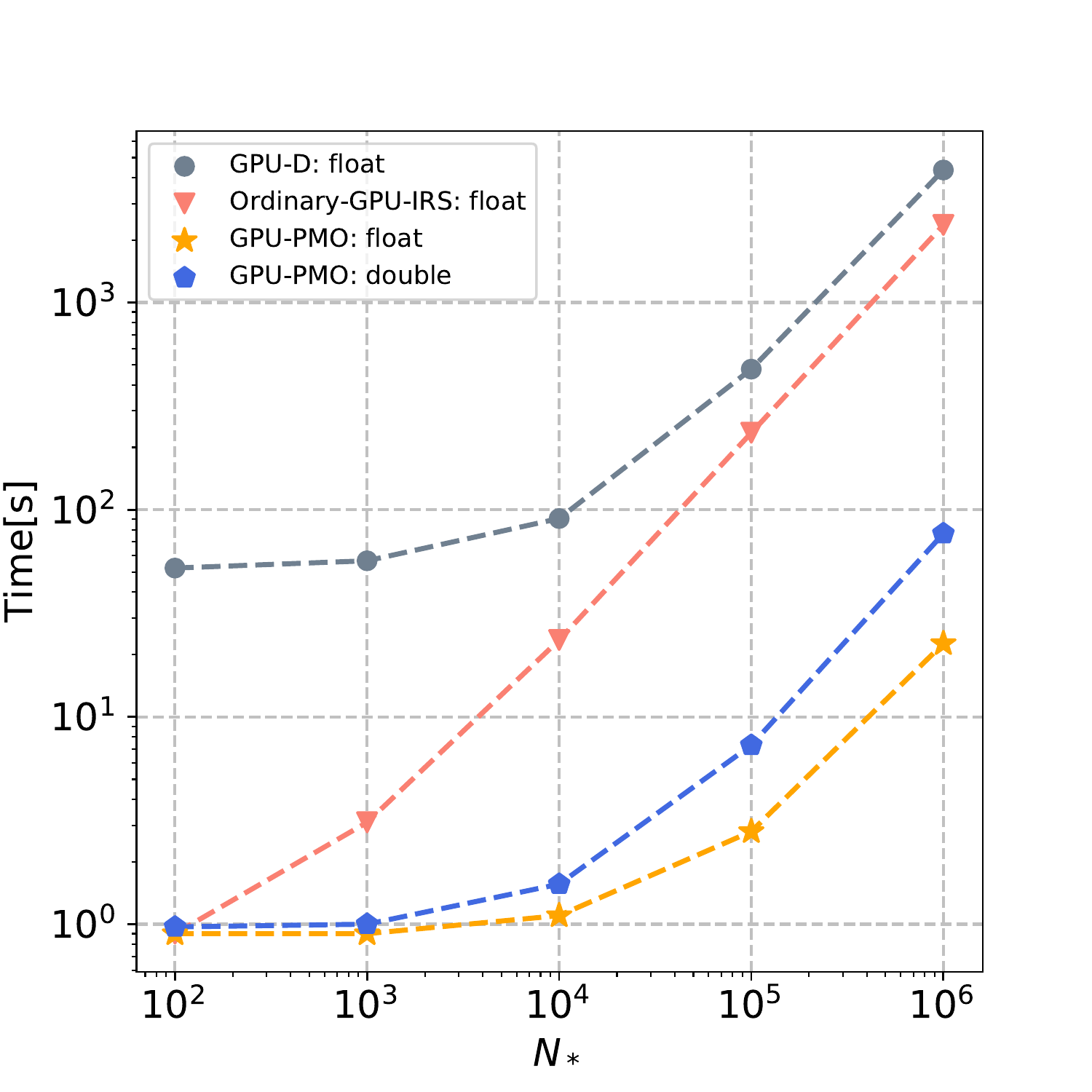}
    \caption{Timing result of our implementation (red triangle for ordinary GPU-based IRS code, yellow five-pointed star for GPU-PMO code in float version, blue pentagon for GPU-PMO code in double version) and GPU-D (gray dot) as a function of $N_*$, varied over the range of $10^2 \sim 10^6$, while $N_{av}$=100 and $N_{pix}=4096^2$ are kept fixed.
    }
    \label{fig:Nstar_speed}
\end{figure}
\;\;\;We download the latest GPU-D source code from the website mentioned in \cite{2014ApJS..211...16V}, \cite{2015ApJS..217...23V} \footnote{https://gerlumph.swin.edu.au}. 
In addition, each performance has been run under the same hardware condition with a GPU card of Nvidia Tesla V100S GPU 32GB.

It should be pointed out that GPU-based programs need to be optimized for specific hardware, however, we cannot ensure that the GPU-D code is optimized to fully fit our hardware. Therefore, we take an IRS program without interpolation method implemented by ourselves as a substitute for the optimized GPU-D.
In particular, we use ``double'' data type in our implementation instead of ``float'' type used in GPU-D, we consider it to be necessary when the image plane is extremely large. However, for a comparison, both ``double'' type and ``float'' type are applied on GPU-PMO code. 
Besides, in our implementation, a rectangular lens plane is used instead of the standard circular one as GPU-D used, therefore, we adjust our realization to fit GPU-D: four codes share the same circular lens plane and random lenses generated by GPU-D code.

\begin{figure}
	\includegraphics[width=\columnwidth]{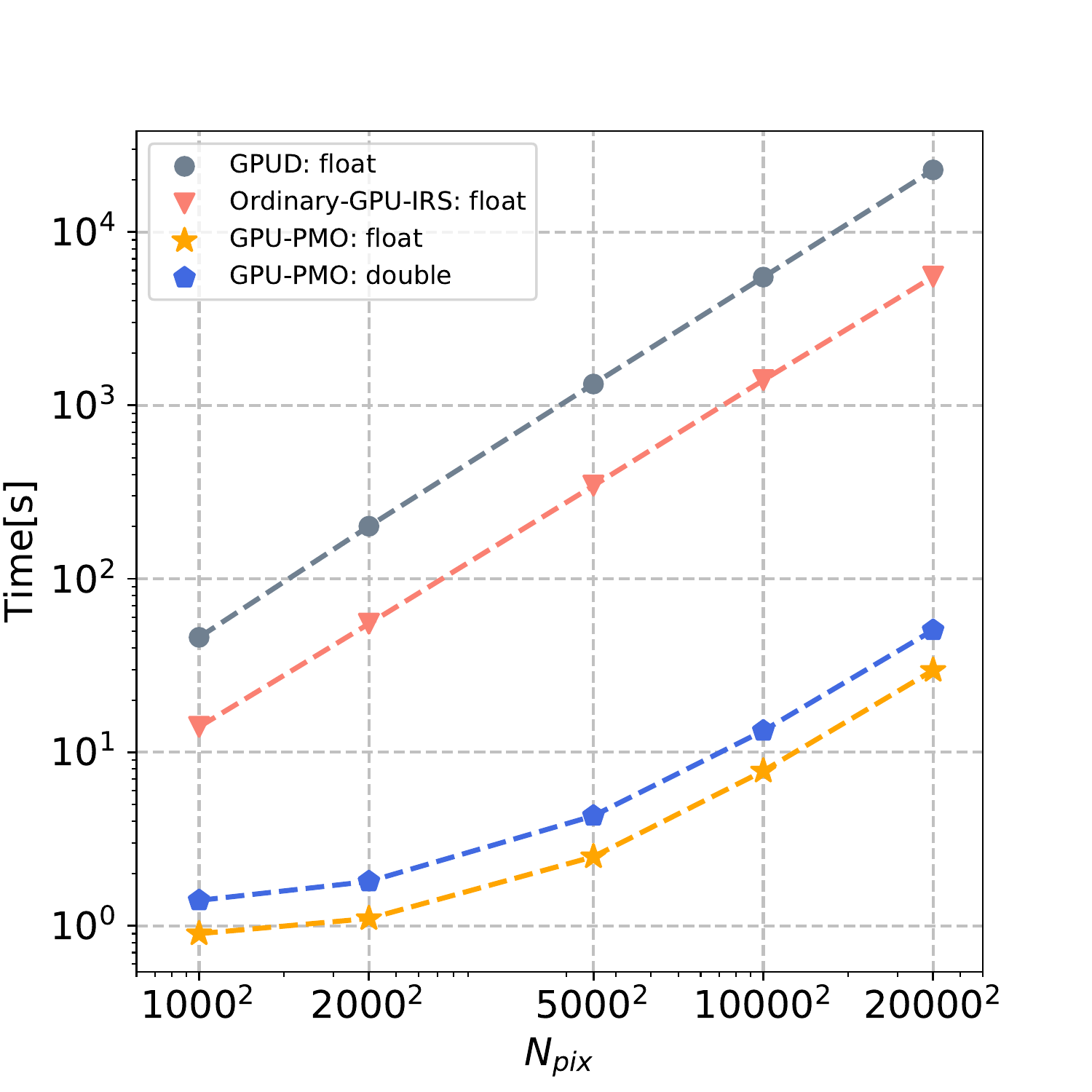}
    \caption{Timing result of our implementation (red triangle for ordinary GPU-based IRS code, yellow five-pointed star for GPU-PMO code in float version, blue pentagon for GPU-PMO in double version) and GPU-D (gray dot) as a function of $N_{pix}$ (corresponding to $1000^2, 2000^2, 5000^2, 10000^2, 20000^2$ ), $N_*$ is fix to $10^4$ and $N_{av}$ = 1000.
    }
    \label{fig:Npix_speed}
\end{figure}

Since run-time T is proportional to the number of $N_{rays}$ and $N_*$:
\begin{equation}
    T \propto N_{av} \cdot N_{pix} \cdot N_* = N_{rays} \cdot N_*,
\end{equation}
we take $N_*$ and $N_{rays}$ as variables: Fig.~\ref{fig:Nstar_speed} and Fig.~\ref{fig:Npix_speed} are the timing result as a function of $N_*$ and $N_{pix}$ (with fixed $N_{av}$), respectively. 
In Fig.~\ref{fig:Nstar_speed}, four programs share the same parameters, with fixed $N_{pix}$ ($4096^2$) and $N_{av}$(100), but a varying $N_*$ from $10^2$ to $10^6$. The advantage of interpolation algorithm becomes apparent when the number of lenses is greater than $10^3$. When $N_*$ is up to $10^5$, GPU-PMO (float type) can be almost 100 times faster than ordinary GPU-IRS and GPU-D.
A similar conclusion can be drawn from Fig.~\ref{fig:Npix_speed}: the time consumption is reduced by about two orders of magnitude when $N_{pix}$ is up to $5000^2$, which is a real boost since 
high resolution maps are required with the development of lensed supernova and other high-precision researches \cite{2018MNRAS.478.5081F}, \cite{2020A&A...644A.162S}.

\subsection{Accuracy}

\begin{figure}
	\includegraphics[width=\columnwidth]{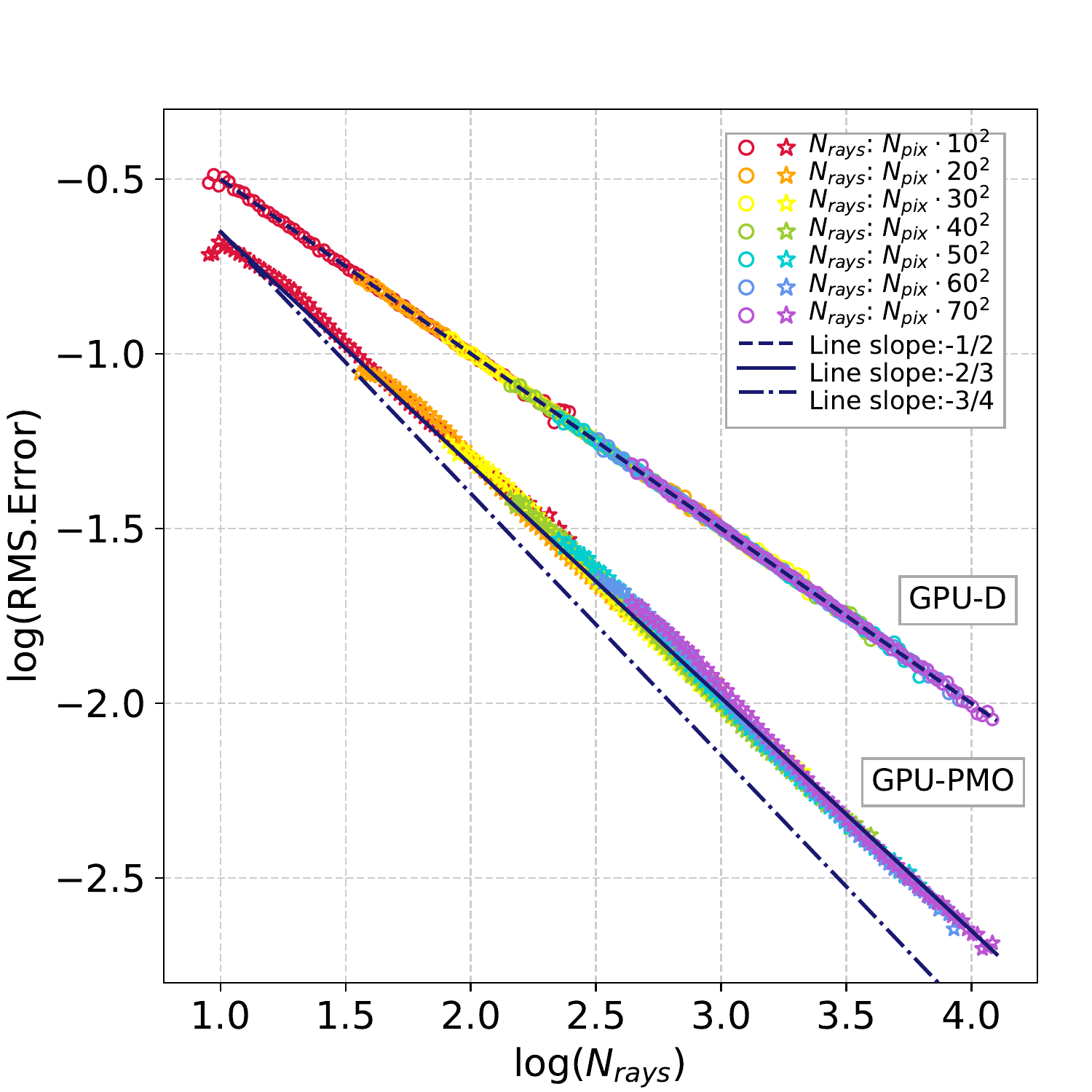}
    \caption{ Accuracy comparison, where $\kappa$=$\kappa_*$=0.652, $\gamma$=0, Nav=100, S\_scale = 50$\theta_E$ with a resolution of $1024^2$, uniform mass distribution with $\left \langle M \right \rangle =0.1M_{\odot}$, corresponding to $10^5$ lenses. Different colors represent different $N_{rays}$, increasing from $10240^2$ to $71680^2$. The blue solid line, blue dotted line and blue dotted and dashed line represent the line with slope of -2/3,-1/2,-3/4 respectively.
    }
    \label{fig:Accuracy}
\end{figure}

We select a set of parameters from the speed comparison of Fig.~\ref{fig:Nstar_speed} to test the accuracy, where $\kappa$ = $\kappa_*$ = 0.652, $\gamma$= 0, $N_{av}$= 100, S\_scale = 50$\theta_E$ with a resolution of $1024^2$, uniform mass distribution with $\left \langle M \right \rangle =0.1M_{\odot}$, corresponding to $10^5$ lenses.

 Due to the nature of IRS method --- the more light rays, the higher the accuracy of magnification map, we use the map generated by ordinary IRS code, which shoots back about 4*$10^{10}$ rays as the criterion.
 Then seven magnification maps with different accuracy (with different number of light rays, from $10240^2$ to $71680^2$) are generated by GPU-PMO and GPU-D respectively, corresponding to seven colors in Fig.~\ref{fig:Accuracy}.
 Thus, we compare the magnification pixel by pixel to measure accuracy: each pixel carries different number of light rays $N_{rays}$, which leads to different magnification distribution. Therefore we arrange 300 logarithmic bins by $N_{rays}$, as the x-axis of Fig.~\ref{fig:Accuracy}, then we obtain the magnification of the pixels and calculate the root mean square (RMS) of the relative error within each bin, the logarithmic values are shown on y-axis.
 
It's clear that for both implementations, the errors are decreasing with the increasing number of light rays. The error of GPU-PMO is smaller at all conditions of $N_{rays}$. Additionally, the larger error of GPU-D may be due to the Poisson error brought by randomly generated light rays, leading to logarithmic error slope close to -1/2 (blue dotted line) as shown in Fig.~\ref{fig:Accuracy}. For ordinary IRS method with uniformly generated light rays, the slope is about -3/4 (blue dashed and dotted line) as described in \cite{2011ApJ...741...42M}.
In our realization we choose to generate the light rays uniformly instead to reduce the error, resulting in a slope around -2/3 (blue solid line), larger then Possion error of -1/2, smaller than theoretical approximation of -3/4.

\section{Summary}\label{sec:Summary}
Cosmological microlensing plays an important role in multi-imaged quasar and even supernova systems, with the upcoming new surveys, increasing number of multi-imaged systems can be expected. 
Extensive simulations for these scenarios can give us a better understanding of lens mass distributions, source inner structures and basic parameters of our universe. 
It implies that large numbers of magnification maps are required to deal with degenerated parameters and obtain statistical results with high confidence. 
Efficient and accurate strategy for generating magnification maps is essential.
In this work, we proposed an optimized GPU-based code to achieve such a goal:

1. We modified the lens equation to use a rectangular lens plane in place of a standard circular one to greatly reduce the unnecessary lenses when handling extreme parameters. In the meantime, we add a ``protective border'' to ensure a guaranteed accuracy of magnification map.

2. We have proposed a GPU-based version of inverse ray-shooting microlensing simulation improved with an interpolation algorithm. 
We achieved this by setting up three-level of grids in the image plane, to divide the lenses into ``nearby lenses'' and ``distant lenses'' for individual light rays, their deflection angles contributed by ``nearby lenses'' will be accumulated precisely. On the other hand, the contribution from ``distant lenses'' will be handled with the level-2 grid points using interpolation strategy, and all the time-consuming part are accomplished on GPU in the mean time.
This approach has greatly reduce the run-time while maintaining accuracy.

3. We have achieved excellent speed and accuracy for generating high-resolution and high-precision magnification maps. When compared with ordinary GPU-IRS and GPU-D,  GPU-PMO code can be about 100 times faster with large number of $N_*$. If dealing with high resolution situation, the time consumption can also be reduced by about two orders of magnitude. Additionally, we have extended our code to multiple-GPU computer in order to achieve better performance when facing up to more challenging numerical situation.

This code is still being polished, you are welcomed to contact us to
obtain the latest version.

\section*{Acknowledgements}
The authors thank Xinzhong Er and Zuhui Fan for helpful discussions and suggestions.
This work is supported by the NSFC (U1931210, No.11673065, 11273061).
We acknowledge the cosmology simulation database (CSD) in the National Basic Science Data Center (NBSDC) and its funds the NBSDC-DB-10 (No. 2020000088)
We acknowledge the science research grants from the China Manned
Space Project with NO.CMS-CSST-2021-A12.

\appendix

\section{Rectangular lens plane correction}
\;\;\;The negative mass sheet  $\pmb\alpha_{-\kappa_*}$ for rectangular lens/image plane:
we obtain the deflection angle of a rectangular plane with smooth mass distribution to a light ray by direct integration,

\begin{equation}
\alpha_{-\kappa_*\_1}=A\cdot \iint \frac{\theta_1-\theta_1^{\prime}}{{(\theta_1-\theta_1^{\prime})}^2+{(\theta_2-\theta_2^{\prime})}^2}\mathrm{d\theta_1^{\prime}}\mathrm{d\theta_2^{\prime}},
\end{equation}

\begin{equation}
A=\frac{\kappa_*}{\pi},  \;\;\;  \mathrm{when} \; \theta_E=1,
\end{equation}

The integral region is the area of the image/lens plane, X-axis direction ($\theta_1^{\prime}$) from $a_1$ to $a_2$, Y ($\theta_2^{\prime}$) from $b_1$ to $b_2$, then: 

\begin{equation}
\begin{aligned}
     \alpha_{-\kappa_*\_1} = \;&A\cdot \int_{b_1}^{b_2}\mathrm{d\theta_2^{\prime}} \int_{a_1}^{a_2}  \frac{\theta_1-\theta_1^{\prime}}{{(\theta_1-\theta_1^{\prime})}^2+{(\theta_2-\theta_2^{\prime})}^2}\mathrm{d\theta_1^{\prime}}\\
              = \;&A\cdot\frac{1}{2}(\theta_2-b_2)\ln_{}{\frac{(\theta_1-a_2)^2+(\theta_2-b_2)^2}{(\theta_1-a_1)^2+(\theta_2-b_2)^2} }\\
              &- A\cdot\frac{1}{2}(\theta_2-b_1)\ln_{}{\frac{(\theta_1-a_2)^2+(\theta_2-b_1)^2}{(\theta_1-a_1)^2+(\theta_2-b_1)^2} } \\
              &+ A\left[(\theta_1-a_2)\arctan\frac{\theta_2-b_2}{\theta_1-a_2} - (\theta_1-a_2)\arctan\frac{\theta_2-b_1}{\theta_1-a_2} \right]\\
              &+ A\left[(\theta_1-a_1)\arctan\frac{\theta_2-b_1}{\theta_1-a_1} - (\theta_1-a_1)\arctan\frac{\theta_2-b_2}{\theta_1-a_1} \right].
\end{aligned}
\end{equation}
\\
Similarly,
\begin{equation}
\begin{aligned}
    \alpha_{-\kappa_*\_2} = \;&A\cdot \iint \frac{\theta_2-\theta_2^{\prime}}{{(\theta_1-\theta_1^{\prime})}^2+{(\theta_2-\theta_2^{\prime})}^2}\mathrm{d\theta_1^{\prime}}\mathrm{d\theta_2^{\prime}}\\
             = \;&A\cdot \int_{a_1}^{a_2}\mathrm{d\theta_1^{\prime}} \int_{b_1}^{b_2}  \frac{\theta_2-\theta_2^{\prime}}{{(\theta_1-\theta_1^{\prime})}^2+{(\theta_2-\theta_2^{\prime})}^2}\mathrm{d\theta_2^{\prime}}\\
             = \;&A\cdot\frac{1}{2}(\theta_1-a_2)\ln_{}{\frac{(\theta_1-a_2)^2+(\theta_2-b_2)^2}{(\theta_1-a_2)^2+(\theta_2-b_1)^2} }\\
             &- A\cdot\frac{1}{2}(\theta_1-a_1)\ln_{}{\frac{(\theta_1-a_1)^2+(\theta_2-b_2)^2}{(\theta_1-a_1)^2+(\theta_2-b_1)^2} } \\
             &+ A\left[(\theta_2-b_2)\arctan\frac{\theta_1-a_2}{\theta_2-b_2} - (\theta_2-b_2)\arctan\frac{\theta_1-a_1}{\theta_2-b_2} \right]\\
             &+ A\left[(\theta_2-b_1)\arctan\frac{\theta_1-a_1}{\theta_2-b_1} - (\theta_2-b_1)\arctan\frac{\theta_1-a_2}{\theta_2-b_1} \right].
\end{aligned}
\end{equation}
\\

Accordingly, the potential for a rectangular area is
\begin{equation}
\begin{aligned}
    \phi_{-\kappa_*}=& \frac{A}{2}\cdot \int_{a_1}^{a_2}\int_{b_1}^{b_2}\ln\left[\left(\theta_1-\theta_1^{\prime}\right)^2+\left(\theta_2-\theta_2^{\prime}\right)^2\right]\mathrm{d\theta_1^{\prime}}\mathrm{d\theta_2^{\prime}}\\
    =& \frac{A}{2}\cdot( \theta_1^{\prime}-\theta_1)^2\mathrm{arctan} \frac{\theta_2^{\prime}-\theta_2}{\theta_1^{\prime}-\theta_1} + ( \theta_2^{\prime}-\theta_2 )^2\mathrm{arctan} \frac{\theta_1^{\prime}-\theta_1}{\theta_2^{\prime}-\theta_2}+(\theta_1^{\prime}-\theta_1)(\theta_2^{\prime}-\theta_2)\ln \left [ (\theta_1^{\prime}-\theta_1)^2+(\theta_2^{\prime}-\theta_2)^2 \right ]\\
    &-3(\theta_1^{\prime}-\theta_1)(\theta_2^{\prime}-\theta_2)\;\;| _{\theta_1^{\prime}=a_1}^{\theta_1^{\prime}=a_2}\;\;| _{\theta_2^{\prime}=b_1}^{\theta_2^{\prime}=b_2},
\end{aligned}
\end{equation}
similar to the equation form in the APPENDIX A of \cite{1998MNRAS.294..734A}. Bring in the upper and lower limits of the integral and we get
\begin{equation}
\begin{aligned}
    \phi_{-\kappa_*}= & \frac{A}{2}\cdot \left \{ \left(\theta_2-b_2\right)\left(\theta_1-a_2\right)\ln\left[\left(\theta_1-a_2\right)^2+\left(\theta_2-b_2\right)^2\right] \right.\\
&-\left(\theta_2-b_2\right)\left(\theta_1-a_1\right)\ln\left[\left(\theta_1-a_1\right)^2+\left(\theta_2-b_2\right)^2\right]\\
&-\left(\theta_2-b_1\right)\left(\theta_1-a_2\right)\ln\left[\left(\theta_1-a_2\right)^2+\left(\theta_2-b_1\right)^2\right]\\
&+\left(\theta_2-b_1\right)\left(\theta_1-a_1\right)\ln\left[\left(\theta_1-a_1\right)^2+\left(\theta_2-b_1\right)^2\right]\\
&+\left(\theta_2-b_2\right)^2\left[\tan^{-1}\frac{\theta_1-a_2}{\theta_2-b_2}-\tan^{-1}\frac{\theta_1-a_1}{\theta_2-b_2}\right]\\
&-\left(\theta_2-b_1\right)^2\left[\tan^{-1}\frac{\theta_1-a_2}{\theta_2-b_1}-\tan^{-1}\frac{\theta_1-a_1}{\theta_2-b_1}\right]\\
&+\left(\theta_1-a_2\right)^2\left[\tan^{-1}\frac{\theta_2-b_2}{\theta_1-a_2}-\tan^{-1}\frac{\theta_2-b_1}{\theta_1-a_2}\right]\\
&-\left(\theta_1-a_1\right)^2\left[\tan^{-1}\frac{\theta_2-b_2}{\theta_1-a_1}-\tan^{-1}\frac{\theta_2-b_1}{\theta_1-a_1}\right]\\
& \left.+3\left(b_2-b_1\right)\left(a_1-a_2\right) \right \}.
\end{aligned}
\end{equation}


\bibliography{References}{}
\bibliographystyle{aasjournal}


\end{sloppypar}
\end{document}